# Theoretical comparison of quantum and thermal noise squeezing in silicon and graphene nanoresonators

Sheping Yan, Yang Xu*, Zhonghe Jin, and Yuelin Wang

*Department of Information Science and Electronic Engineering, Zhejiang University, Hangzhou, China, 310027*

We theoretically compared quantum noise squeezing differences between silicon and graphene nanoresonators based on experimental structure parameters. The conditions to achieve squeezed states of silicon and graphene have been discussed. According to our theoretical analysis, graphene nanoresonators can obtain a much smaller squeezing factor than silicon, taking advantage of their thin thickness. Both the quantum noise and thermal noise (Brownian motion) of typical monolayer graphene nanoresonator can be reduced by 12.58 dB at $T$ = 5 K with a pump voltage of 5 V.

The Heisenberg uncertainty principle imposes an intrinsic limitation,[1,2] known as quantum noise, on the ultimate sensitivity of quantum measurement systems such as atomic forces,[3] infinitesimal displacement,[4] and gravitational-wave detections.[5] Quantum squeezing is an efficient way to decrease the system quantum noise and develop a high precision detection mechanism.[6,7] Thermomechanical noise squeezing has been studied by D. Rugar *et al*,[8] where the resonator motion in the fundamental mode was parametrically squeezed in one quadrature by periodically modulating the effective spring constant at twice the frequency. Quantum noise squeezing has been theoretically analyzed in silicon (Si) nanoresonators with rigorous conditions at 20 mK with a pump voltage of 5 V.[9,10] Latterly, Junho Suh *et al*.[11] have successfully achieved parametric amplification and back-action noise squeezing using a qubit-coupled Si nanoresonator. Recently, free-standing graphene membranes have been fabricated,[12] which have high carrier mobility $\mu$,[13] small volumetric mass density, stable monolayer structure,[14] providing an excellent platform to develop transducers with high sensitivity. The magnitude of quantum and thermal noise in graphene nanoelectromechanical system (NEMS) is comparable to that of signals caused by external molecular adsorbates or electrostatic forces. However, little attention is devoted to the quantum and thermal noise squeezing in graphene NEMS. In this work, we find that achieving squeezed states is crucial to improve the ultimate precision of graphene-based transducers.

We start with a theoretical comparison of the quantum noise squeezing in Si[10] and graphene[15] nanoresonators by analyzing the experimental devices. To study squeezing effects in mechanical systems, zero-point displacement uncertainty $\Delta x_{zp}$ caused by intrinsic quantum noise or is introduced. In classical mechanics, the complex amplitudes of an oscillator or resonator $X = X_1 + iX_2$, where $X_1$ and $X_2$ are the real and imaginary parts of complex amplitudes respectively, can be obtained with complete precision. In quantum mechanics, $X_1$ and $X_2$ do not commute and satisfy the uncertainty relationship $\Delta X_1 \Delta X_2 \geq \hbar/2M_{eff}w$, where $\Delta X_1$ and $\Delta X_2$ are the quantum fluctuations of $X_1$ and $X_2$, respectively,[6] $\hbar$ is the reduced Planck constant, $M_{eff} = M/4 = \rho LWh/4$ is the film's effective motional mass,[10] $M$ is the physical mass, $\rho$ is the volumetric mass density, $L$, $W$, and $h$ are the length, width, and thickness of the film, respectively, and $w$ is the fundamental flexural mode angular frequency with $w = 3.52hE^{1/2}\rho^{-1/2}L^{-2}$ for a cantilever and $w = 6.48hE^{1/2}\rho^{-1/2}L^{-2}$ for a double-clamped film, and $E$ is the Young's modulus of the material. Then, intrinsic quantum noise $\Delta x_{zp}$ of the nanoresonantor can be deduced to $\Delta x_{zp} = \Delta X_1 = \Delta X_2 = [\hbar/(2M_{eff}w)]^{1/2}$.[1]

A typical nanoresonator with a double-clamped free-standing nanofilm is schematically shown in Fig. 1. The substrate is doped Si with high conductivity and the SiO$_2$ insulator layer is used to separate the substrate and the film. A pump voltage is applied between the film and the substrate. All of the structural and material parameters in our simulation are adopted from the experimental data of the corresponding devices.[15] For graphene film, $E_{graphene}$ = 1.03 × 10$^{12}$ Pa, $\rho_{graphene}$ = 2.21 Kg/m$^3$, and $h_{graphene}$ = 0.335 nm for monolayer, $h_{graphene}$ = 0.670 nm for bilayer, and $h_{graphene}$ = 1.005 nm for trilayer based on previous theories and scanning tunneling microscope experiments.[16,17,18] For a Si beam, $E_{si}$ = 1.50 × 10$^{11}$ Pa, $\rho_{si}$ = 2.33 × 10$^3$ Kg/m$^3$, and $h_{si}$ = 0.1μm.[10] $\Delta x_{zp}$ of Si beams and multilayer graphene films with various dimensions are compared in Fig. 2(a) and (b), respectively. $\Delta x_{zp}$ of graphene can be several angstroms and is four orders of magnitude larger than that of silicon for similar geometrical sizes. The 0.33 nm thickness of monolayer graphene imposes a much larger quantum fluctuation than Si nanofilm, which is usually in tens of nanometers thick. Therefore, quantum noise squeezing is more important in graphene NEMS than Si NEMS to improve signal-to-noise ratio.

To analyze quantum noise squeezing in nanoresonators, a parametric pumping is applied at the frequency $2w$ to modulate the effective spring constant $k$ of the membrane film.[6,8,10] Two assumptions are used, namely, the film width $W$ is in micrometer scale and $|X_1| \ll d$, where $d$ is the distance between the film and the substrate. Applying a pump voltage $V_a(t) = V\sin(2w_a t + \theta)$, between the membrane film and the substrate, the spring constant $k$ will have a sinusoidal modulation $k_a(t)$,[8,10] given by $k_a(t) = -\sin(2w_a t + \theta)\Delta k$, where $\Delta k = C_T V^2/2d^2$ and $C_T$ is the total capacitance composed of structure capacitance $C_0$, quantum capacitance $C_q$, and screen capacitance $C_s$ in series.[19] $C_q$ and $C_s$ of graphene can not be neglected [20,21] owing to the thickness of graphene film at atomic scale, while those capacitances of silicon are usually negligible which are also discussed and confirmed in Ref. 8. The quantum capacitance of monolayer graphene [22] is $C_{q(monolayer)} = 2e^2 n^{1/2}/(\hbar v_F \pi^{1/2})$, where $n$ is the carrier concentration, $v_F \approx c/300$ and c is the velocity of light, $C_q$ $C_q = 2m^* e^2/\pi\hbar^2$ with $m^*=0.037 m_e$ for bilayer and $m^*=0.053 m_e$ for trilayer graphene,[23] where $m_e$ is the electron mass. By changing $\theta$, different spring modulations can be obtained, shown in Fig. 3. Pumping the suspended nanofilm from an initial thermal equilibrium state at frequency $w_a = w$, the displacement uncertainty in $X_{1,2}(t)$, $\Delta X_{1,2}(t, \theta)$, can be obtained,[6] expressed as

$$\Delta X_{1,2}(t,\theta) = (\hbar/2M_{eff}w)(2N+1)e^{-t/\tau}[ch(\alpha t) \mp \cos\theta sh(\alpha t) + \tau^{-1}(I_a \pm \cos\theta I_b)] \qquad (1)$$

where $I_a = \int_0^t e^{t/\tau} ch[\alpha(\xi - t)]d\xi$, $I_b = \int_0^t e^{t/\tau} sh[\alpha(\xi - t)]d\xi$, $N = [\exp(\hbar w/k_B T) - 1]^{-1}$ is the average number of quanta at absolute temperature $T$ and frequency $w$, $k_B$ is the Boltzmann constant, $\tau = Q/w$ is the relaxation time of the mechanical vibration, where $Q$ is the quality factor of the nanofilm resonator and $\alpha = \Delta k/2M_{eff}w$.

As shown in Fig. 3, when $\theta = 0$, a maximum modulation state in $X_1$ can be reached,[6,8] and $\Delta X_1$ can be simplified as $\Delta X_1(t) = (\hbar/2M_{eff}w)(2N+1)(\tau^{-1}+\alpha)^{-1}(\tau^{-1}+\alpha\exp[-(\tau^{-1}+\alpha)t])$. As $t \to \infty$, the maximum squeezing of $\Delta X_1$ is always finite, with expressions of $\Delta X_1(t \to \infty) \approx [\hbar(2N+1)(1+QC_T V^2/(4d^2 M_{eff} w^2))^{-1}/2M_{eff}w]^{1/2}$. The squeezing factor $R$, defined as $R = \Delta X_1/\Delta x_{zp} = \Delta X_1/(\hbar/2M_{eff}w)^{1/2}$, can be written as follows when $\theta = 0$,

$$R = \sqrt{\frac{2/(\exp[\hbar(k_B T)^{-1}(w^2 + E\varepsilon\pi^2 \rho'^{-1} L^{-2})^{1/2}]-1)+1}{1+QV^2[2d^2(1/C_0 + 1/C_q + 1/C_s)]^{-1}[2M_{eff}(w^2 + E\varepsilon\pi^2 \rho'^{-1} L^{-2})^{1/2}]^{-1}}} \qquad (2)$$

where $\varepsilon$ is an applied strain on the film, $\rho'$ represents the effective volumetric mass density of film after applying strain. In order to achieve quantum squeezing, $R$ must be less than 1. $R$ values of various silicon and graphene films with applied voltages at $T = 5$ K are shown in Fig. 4. Squeezing factors $R$ of multilayer graphene films follow the relation $R_{trilayer} > R_{bilayer} > R_{monolayer}$, shown in Fig. 4(b), as $R$ is proportional to the thickness of the graphene film. The thicker the film, the more difficult it is to achieve a quantum squeezed state, which also explains why traditional Si NEMS are not easy to achieve quantum squeezing due to their usual thickness at several hundred nanometers, shown in Fig. 4(a). $R$ approaches unity as $L$ approaches zero, while $R$ tends to be zero as $L$ approaches infinity. It explains why $R$ has some kinked regions, shown in the upper right part of Fig. 4(b) with black circle, where the graphene film length is at nanometer scale.

By choosing a typical monolayer graphene nanoresonator in Ref. 13 with $L = 1.8$ μm, $W = 0.2$ μm, $d = 0.1$ μm, $T = 5$ K, $Q = 14000$, and $V = 500$ mV, we obtained $\Delta x_{zp(graphene)} = 0.0416$ nm, $R_{graphene} \approx 0.2394$, and after applying quantum squeezing effects $\Delta x_{zp}$ can be reduced to 0.01 nm, a 12.53-dB reduction relative to the quantum noise. While, for a typical Si nanoresonator, $\Delta x_{zp(Si)} = 3.9594 \times 10^{-5}$ nm, $R_{Si} = 28.4172$. The squeezing factor of Si nanoresonator can only reach $R_{Si} \approx 0.42$ even at $T = 20$ mK and $V = 5$ V,[10] in contrast, the squeezing factor of a typical graphene NEMS in Ref. 13 is as small as $R_{graphene} \approx 0.0016$ at the same environmental condition as shown in Fig. 4, which means a 55.92 dB reduction relative to the quantum noise. We can conclude that quantum noise squeezing of graphene nanoresonators can be achievable with $T = 5$ K, $V = 500$ mV, while silicon can not achieve quantum squeezing unless increasing the applied pumping voltages and reducing the temperature. Compared with the traditional Si nanoresonators, the graphene film provides a larger room to satisfy quantum squeezing conditions.

In contrast to the formal squeezing analysis proposed by D. Rugar *et al*,[8] in which steady-state solutions have been assumed and the minimum $R$ is 1/2, we use time dependent pumping techniques to prevent $X_2$ from growing without bound as t → ∞, which should be terminated after the characteristic time $t_c = \ln(Q\Delta k/2M_{eff}w^2)2M_{eff}w^2/\Delta k$, when $R$ achieves its limiting value. Therefore, we have no upper bound on $R$. We have to point out that this method needs a precise phase control of $\Delta\theta < 2\times 10^{-7}$ shown in Fig. 3. To make the heating from mechanical energy conversion negligible during the pump stage, $t_c \ll \tau$ must be satisfied. We find $t_{c(graphene)}$ = 1.86 ns, $\tau_{(graphene)}$ = 30.51 μs, and $t_{c(graphene)}/\tau_{(graphene)} \approx 6.11 \times 10^{-5}$ for the monolayer graphene parameters considered in the text, while $t_{c(si)}$ = 17.74 μs, $\tau_{(si)}$ = 49.06 μs, $t_{c(si)}/\tau_{(si)} \approx 0.36$ for the Si parameters with the same environmental conditions, and $t_{c(si)}/\tau_{(si)} \approx 0.03$ when $V$ = 5 V and $T$ = 20 mK. Fig. 5 has shown the time dependence of $\Delta X_{1,2}$ in typical silicon and monolayer graphene nanoresonators with $T$ = 5 K and $V$ = 500 mV, when $\theta$ = 0. According to Fig. 5(b), the squeezing has reached the limiting value after one $t_c$ time.

However, to measure the quantum noise, the intrinsic Brownian motion noise of graphene nanoresonator also needs to be considered. The graphene film thermal vibration can be estimated from Treacy *et al*.[24] According to the model, the vibration amplitude of a double-clamped graphene film is $x^2_b = \Sigma k_B T/k_n$, where $k_n = \beta_n^4 EWh^3/12L^3$ is the effective spring constant for the middle part motion of nanofilm and the values of $\beta_n$ are the solutions to the equation of $\cos\beta\cosh\beta + 1 = 0$. $\beta_0 = 0$, $\beta_1 \approx 4.730$, $\beta_2 \approx 7.8532$, $\beta_3 \approx 10.996$, and $\beta_n \approx (n + 1/2)\pi$. Then, $x_b = 12L^3 k_B T/(EWh^3)\Sigma\beta_n^{-4} \approx [0.02857 k_B T L^3/(EWh^3)]^{1/2}$ can be obtained. For a monolayer graphene film considered in this work, we can have $x_b \approx 0.5449$ nm at $T$ = 5 K and $x_b \approx 0.03446$ nm at $T$ = 20 mK. The thermal noise can also be squeezed based on above theoretical analysis, a 12.53-dB reduction relative to the thermal equilibrium value at $T$ = 5 K, and the noise squeezing can be reached to -55.92 dB at $T$ = 20 mK.

To summarize, we presented systematic studies of quantum squeezing effects in silicon and graphene devices as a function of time $t$, applied voltage $V$, and phase $\theta$. We compared theoretical necessary conditions to achieving quantum squeezed states of graphene and silicon nanoresonators.

We gratefully acknowledge Prof. Raphael Tsu at UNCC, Prof. Jean-Pierre Leburton at UIUC, Prof. Yuanbo Zhang at Fudan Univ., Prof. Jack Luo at Univ. of Bolton, and Prof. Bin Yu at SUNY for fruitful discussions and comments. This work is supported by the National Science Foundation of China (Grant No. 61006077) and the National Basic Research Program of China (Grant No.2006CB300405). Dr. Y. Xu is also supported by the Excellent Young Faculty Awards Program (Zijin Plan) at Zhejiang University and Specialized Research Fund for the Doctoral Program of Higher Education (SRFDP with Grant No. 20100101120045).

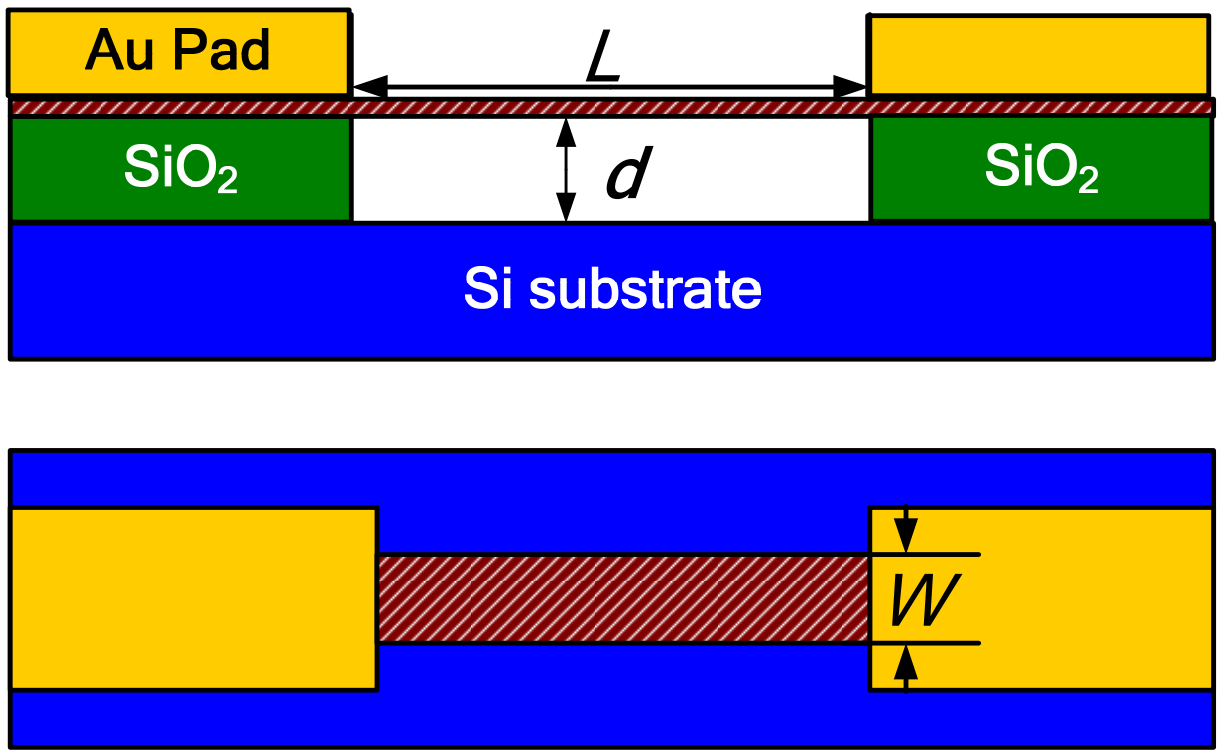

FIG. 1. (color online) Schematic of a suspended graphene or silicon NEMS device.

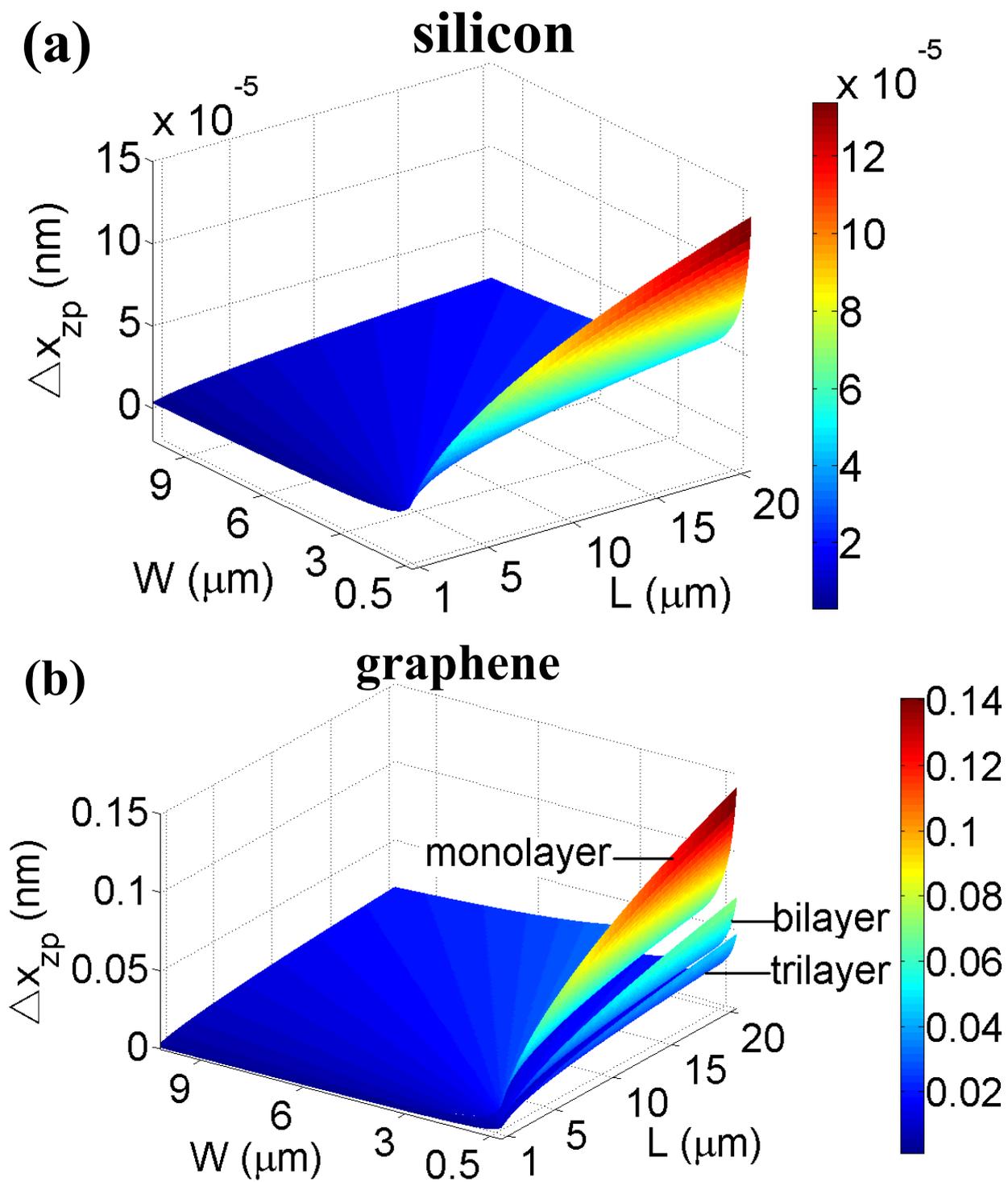

**FIG. 2**. (color online) (a) $\Delta x_{zp}$ versus various silicon beam sizes. (b) $\Delta x_{zp}$ versus various graphene film sizes.

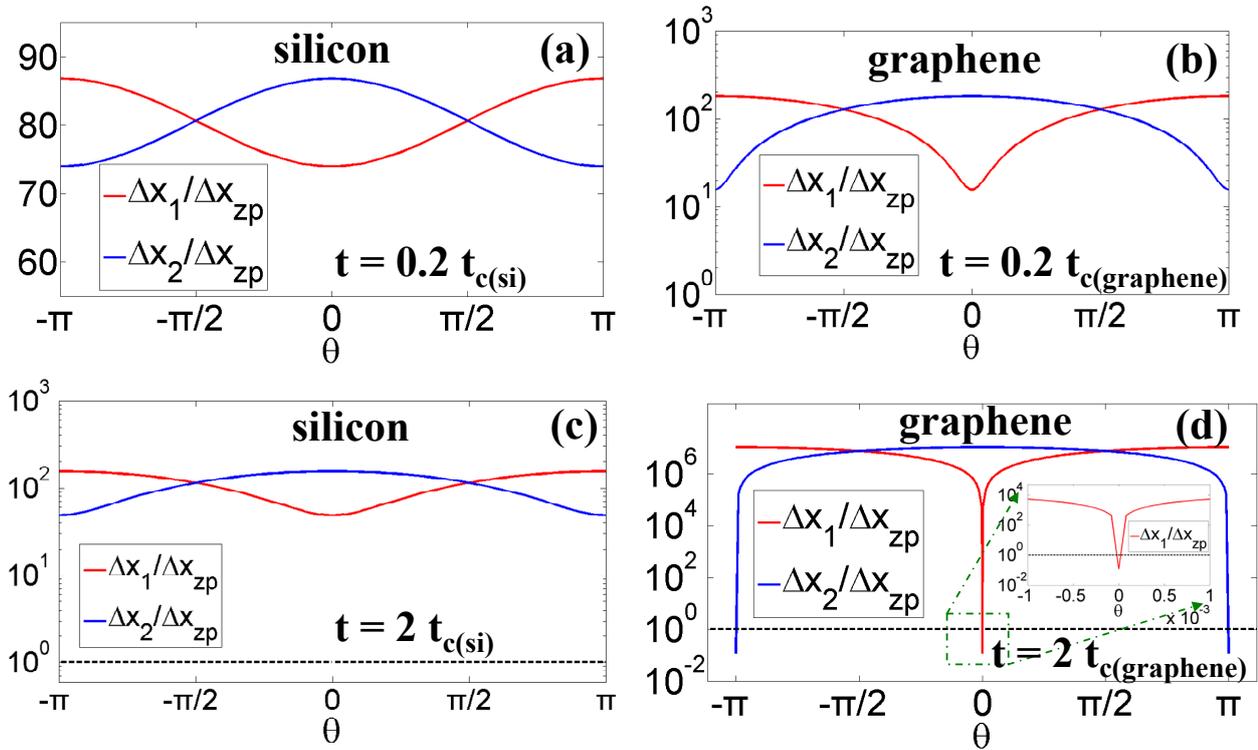

**FIG. 3**. (color online) Phase dependence of $\Delta X_1$ and $\Delta X_2$, which are expressed in units of $\Delta x_{zp}$ with $T = 5$ K, and $V = 500$ mV. The reference line is $\Delta X = \Delta x_{zp}$. (a) silicon at $t = 0.2\, t_{c(si)}$ and (b) monolayer graphene at $t = 0.2\, t_{c(graphene)}$, (c) silicon at $t = 2\, t_{c(si)}$ and (d) monolayer graphene at $t = 2\, t_{c(graphene)}$, where the minimum value of R is 0.2394. The inset shows the precise phase control is required to obtain R<1.

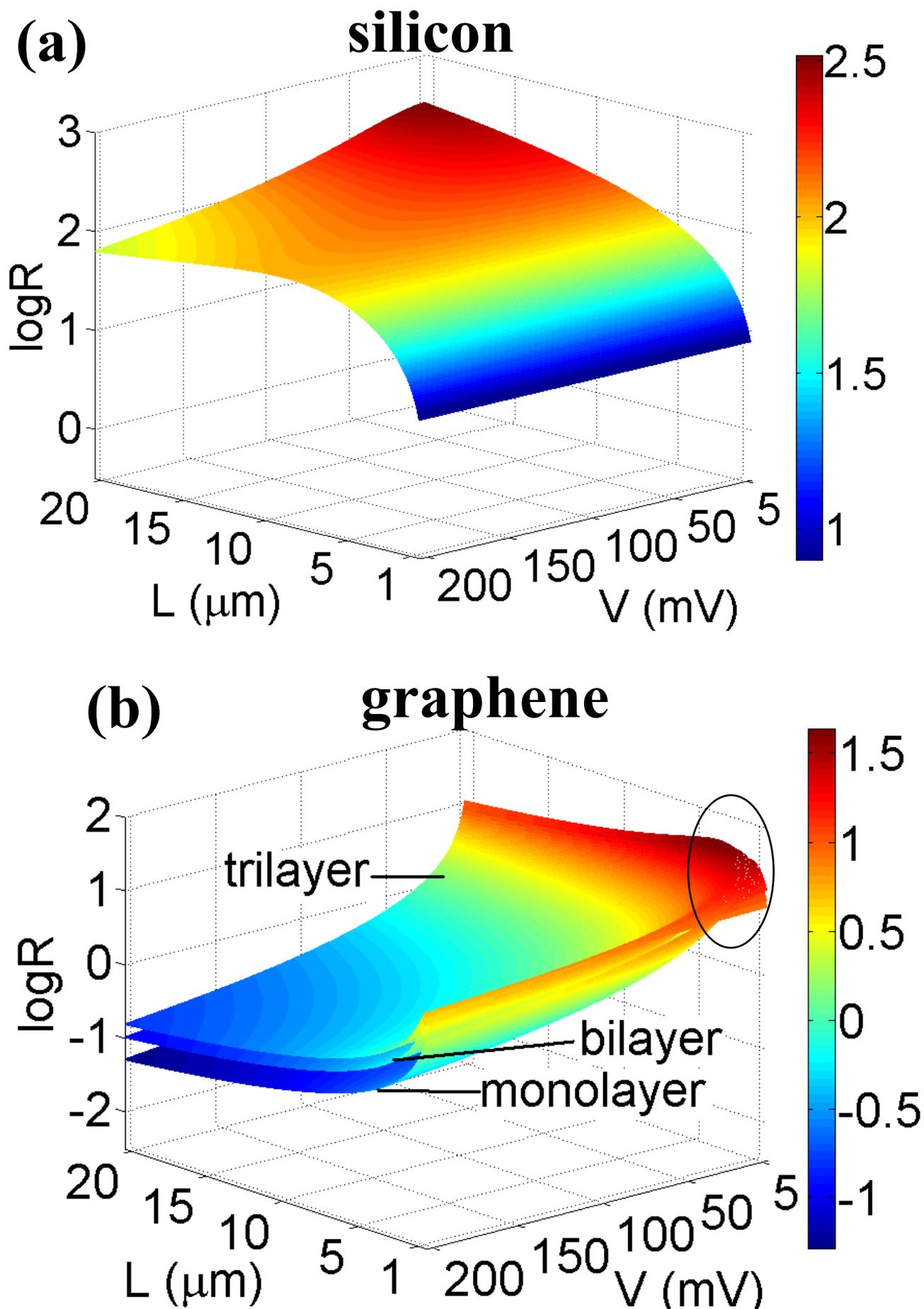

**FIG. 4**. (color online) (a) log$R$ versus various silicon beam lengths and applied voltages at $T$ = 5 K. (b) log$R$ versus various graphene film lengths and applied voltages at $T$ = 5 K.

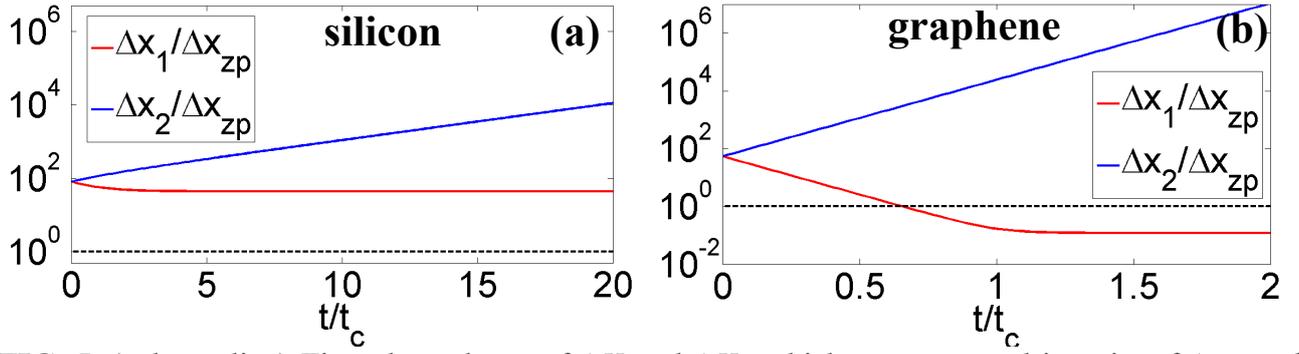

**FIG. 5**. (color online) Time dependence of $\Delta X_1$ and $\Delta X_2$, which are expressed in units of $\Delta x_{zp}$ and time in units of $t_c$, with $T = 5$ K, and $V = 500$ mV when $\theta = 0$, (a) silicon and (b) monolayer graphene. The dashed reference line is $\Delta X = \Delta x_{zp}$.